\documentclass[twocolumn,showpacs,amsmath,amssymb]{revtex4}
\usepackage{graphicx}
\usepackage{color}

\begin{document}
\title{Interlayer Exchange Coupling Beyond the Proximity Force Approximation}
\author{Ching-Hao Chang and Tzay-Ming Hong}
\affiliation{Department of Physics, National Tsing Hua University, Hsinchu, Taiwan 300, Republic of China}
\date{\today}

\begin{abstract}
Ion bombardment has been shown to be capable of enhancing the interlayer exchange coupling in a trilayer system that exhibits giant magnetoresistance.
We demonstrate that this phenomenon can be derived from the
phase coherence
among scattered paths within the two rough interfaces when their topographies are correlated.
In the case of mild corrugations, our method reproduces the predictions by the proximity force approximation which does not consider the interference. When the characteristic Fourier conjugate of the tomography becomes large and comparable to the Fermi momentum, interesting new features arise and can only be captured by our more general approach.
Among our findings, the scenario of an enhanced interlayer exchange coupling due to the interface roughness is explained, along with how it depends on the sample parameters.
An additional channel for the resonant
transmission is  identified due to extra scattering paths from the roughness.
\end{abstract}
\maketitle
\section{Introduction}
Interlayer exchange coupling (IXC) has been studied for more than twenty years\cite{I1,I2,I3,MD,Kudrnovsky,Bruno}
with applications in phenomena such as the Giant Magnetoresistance\cite{g1,g2} (GMR) and the Tunneling Magnetoresistance (TMR).
Due to the lack of reliable microscopic theories, the interface roughness (IR) was mostly treated by the static average\cite{MD,JO,JU} which invariably led to a suppression on IXC. Improvement has been achieved by a systematic study using the perturbation method\cite{pre-art}. We shall follow up this line of approach with more detailed calculations and report new information on how to raise the sensitivity of GMR and why an enhancement in coupling is possible, as has already been observed in ion-bombarded samples\cite{bomb}.

Besides IXC, the Casimir effect between metallic mirrors\cite{Casimir} faces the same complexity because a lot of experiments were performed by using a spherical test body in addition to the unavoidable corrugations on its surface.
This Casimir force has been measured with high precision which
provides a fertile ground to test the theoretical models of IR.
The proximity-force approximation (PFA), equivalent to statically averaging over the plane-plane geometry,
is the first intuitive theory to be examined by both experiments\cite{Casimir-exp1,Casimir-exp2} and theories\cite{pfa-the}.
Since the PFA was shown to work only for mild corrugations\cite{pfa-the},
Maia Neto \textit{et al.} generalized it by the perturbation theory\cite{Casimir-the,uncor-the,cor-the} to obtain better agreements with the experiments\cite{exp-the1,exp-the2}. 
Recently, a series of experiments on severe corrugated mirrors, which are outside of the
applicability regime of the PFA, confirm that a scattering
approach such as Maia Neto's is needed to capture the essence of the nontrivial
diffraction effects.\cite{exp-the3,exp-the4}

We observe that IXC shares the same physical concept and mathematical construction as
the Casimir effect, which connection is proved in Appendix A. A quick way to convince oneself of this similarity is by the expression of the Casimir energy\cite{lossy-cavity-casimir,newj} $E$ between two parallel flat mirrors separated
by a distance $D$ and with area $A$ and reflection coefficient
$r(k_{\|},E)$
\begin{equation}
\frac{E}{A}=\sum_{p}\int_{0}^{\infty} \frac{d\xi}{2\pi} \int \frac{d^{2}k_{\|}}{(2\pi)^{2}}
\mathrm{ln}\Big[1-r^{2}(p)\ e^{-2\sqrt{\frac{\xi^{2}}{c^{2}}+k^{2}_{\|}}D}\Big].
\label{casimir}
\end{equation}
where $\xi$ is the imaginary frequency and $p$ denotes the transverse electric and magnetic modes.
This bears great resemblance to the IXC energy in a trilayer system with two metallic layers and a metallic or insulating spacer\cite{MD}
\begin{equation}
\frac{\triangle E}{A}=2\mathrm{Im}\int_{-\infty}^{E_{F}} \frac{dE}{2\pi} \int \frac{d^{2}k_{\|}}{(2\pi)^{2}}
\mathrm{ln}\Big[1-r^{2}e^{2i\sqrt{2m(E-V_0)-k^{2}_{\|}}D}\Big]
\label{IXC}
\end{equation}
where $k_{\|}$ integration is over the interface Brillouin zone while the $D$ now represents the spacer width.
One purpose of this work is to generalize this comparison to include the effects of IR.
However, the physics in trilayers is more versatile because the spacer exhibits a characteristic length scale, the Fermi wavelength $1/k_F$, compared to which other length scales can be tuned to give different behavior. These parameters include the
roughness amplitude, corrugation length, and spacer width. Furthermore, the fact that the spacer can be either metallic or insulating also
enriches the phenomenon caused by IR.

We apply the perturbation method to calculate IXC in Section II, and demonstrate that the predictions are equivalent to those by PFA
in the limit of smooth corrugations in Section III.
In Section IV, the two interface tomographies in TMR and GMR are assumed to correlated, with special attention to the interference effect on IXC.
Section V is devoted to study why, when and how much the
IXC can be enhanced by IR. Discussions and conclusions are arranged in the final Section VI, where improvements over our previous work are explained. To
preserve the conciseness of the main text, a rigorous proof of the connection between
IXC and the Casimir effect is arranged in Appendix A.

\section{Perturbation method}
In this section, we start by deriving the two-dimensional scattering states
generated by a right-moving plane wave with momentum $(k_{x},k_{y})$ that interacts with an irregular interface $A(y)$ at $x=0$.
The potential on the left side of $A$ is set to be higher in energy by $V_{0}$.
The wavefunctions on the left and right sides are denoted by $\Phi(x,y)$ and $\Psi(x,y)$, respectively.
The boundary conditions are
\begin{align}
\notag
&\Phi(A(y),y) = \Psi(A(y),y)  \\
&\frac{\partial\Phi(x,y)}{\partial x}\big|_{x=A(y)}=\frac{\partial\Psi(x,y)}{\partial x}\big|_{x=A(y)}.
\label{eq:b}
\end{align}

In this work we shall assume that $A$ is much smaller than both $1/k_x$ and the major Fourier corrugation wavelength $\lambda_c$
in order to proceed with the perturbative calculations as in the Casimir effect\cite{uncor-the}.
The scattering states can be obtained by treating the IR as a perturbation to the smooth interfaces,
\begin{align}
\notag
& \Phi(x,y)=\Phi_{0}(x,y)+\sum_{n,q_{y}}a^{(n)}_{k_{y},q_{y}}e^{-iq_{x}x+iq_{y}y}\\
& \Psi(x,y)=\Psi_{0}(x,y)+\sum_{n,q_{y}}b^{(n)}_{k_{y},q_{y}}e^{iq_{x}^{'}x+iq_{y}y}
 \label{p:wave}
\end{align}
where $\Phi_{0}(x,y)$ and $\Psi_{0}(x,y)$ are the unperturbed scattering states, and the transmitted wave $\Psi_{0}(x,y)$ carries momentum $(k'_{x},k_{y})$
and superscript $(n)$ denotes the $n$-th order perturbation.
For an elastic scattering, the dispersion relation in Eq.(\ref{p:wave}) is
\begin{align}
E&=\frac{k^{2}_{x}+k^{2}_{y}}{2m^{*}}+V_{0}=\frac{k^{'2}_{x}+k^{2}_{y}}{2m^{*}}\\
&=\frac{q^{2}_{x}+q^{2}_{y}}{2m^{*}}+V_{0}=\frac{q^{'2}_{x}+q^{2}_{y}}{2m^{*}}
\end{align}
where $m^{*}$ represents the effective mass of the carrier.
Insert Eq.(\ref{p:wave}) into Eq.(\ref{eq:b}) and use
$|k_{x}A(y)|$ and $|k'_{x}A(y)|$ as the perturbation factors to expand Eq.(\ref{eq:b}).
Retaining up to the second order, one can show that\cite{pre-art}
\begin{align}
a^{(1)}_{k_{y},q_{y}}=&-i(q'_{x}-q_{x})t_{k_{x},k'_{x}}\langle  k_{y}|A(y)|q_{y}\rangle \label{eq:a1}\\
\notag a^{(2)}_{k_{y},q_{y}}=& m^{*}V_{0}\frac{k'_{x}+q'_{x}}{q_{x}+q^{'}_{x}}t_{k_{x},k'_{x}}\langle  k_{y}|A^{2}(y)|q_{y}\rangle \\
-&2i\frac{m^{*}V_{0}}{q_{x}+q^{'}_{x}}\sum_{q_{y2}}\langle  q_{y2}|A(y)a^{(1)}_{k_{y},q_{y2}}|q_{y}\rangle
\label{eq:a1a2}
\end{align}
where the subscripts $k_{y},q_{y}$ denotes scatterings from $k_{y}$ to $q_{y}$ state,
and $t_{k_{x},k'_{x}}$ is the transmission coefficient for a smooth interface.

Interlayer exchange coupling in a trilayer system, affected by the quantum interference among the reflected waves,
can be described by the reflection matrices\cite{Bruno}
\begin{align}
\frac{\triangle E}{W}=\mathrm{Im}\int_{-\infty}^{E_{F}}\frac{dE}{2\pi^{2}}
\mathrm{Tr} \big[\mathrm{ln}(I-\hat{R}_{L}^{-+}e^{i\hat{K}^{+}D}\hat{R}_{R}^{+-}e^{i\hat{K}^{-}D})\big],
\label{eq:ixc}
\end{align}
where $W$ is the length of the interface, $I$ is the unit matrix, $\hat{R}_{L}^{-+}/\hat{R}_{R}^{+-}$ are the reflection matrices
from the left/right smooth interfaces.


When the topography $A_{L}(y)/A_{R}(y)$ at the left/right interface is considered,
the reflection matrix can be written in powers of the perturbation:
\begin{equation}
\hat{R}_{L}^{-+}\approx\hat{R}_{L}^{(0)-+}+\hat{R}_{L}^{(1)-+}+\hat{R}_{L}^{(2)-+}.
\end{equation}
The zero-order matrix $\hat{R}_{L}^{(0)-+}$ corresponds to a smooth interface and is diagonal in the basis:

\begin{equation}
\hat{R}_{L}^{(0)-+} = \left[
  \begin{array}{cccc}
    r_{L;k_{x},k'_{Lx}} & 0 & 0 & \cdots \\
    0 & \ddots &  &  \\
    0 &  & r_{L,;q_{x},q'_{Lx}} &  \\
    \vdots &  &  & \ddots \\
  \end{array}
\right],
\end{equation}
where $r_{L;k_{x},k'_{Lx}}/r_{L;q_{x},q'_{Lx}}$ are the reflection coefficients of momenta $k_x/q_x$
from the left interface while $k'_{Lx}/q'_{Lx}$ denotes the momentum in the left side layer. Same for the definition of $\hat{R}_{R}^{+-}$.

The first and second-order matrices is constructed by Eq.(\ref{eq:a1}) and Eq.(\ref{eq:a1a2}) as
\begin{equation}
\hat{R}_{L}^{(n)-+}= \left[
  \begin{array}{cccc}
    a^{(n)}_{L;k_{y},k_{y}} & \cdots & a^{(n)}_{L;q_{y},k_{y}} & \cdots \\
    \vdots & \ddots &  &  \\
    a^{(n)}_{L;k_{y},q_{y}} &  & a^{(n)}_{L;q_{y},q_{y}} &  \\
    \vdots &  &  & \ddots \\
  \end{array}
\right].
\label{eq:r-matrix}
\end{equation}
Inserting Eq.(\ref{eq:r-matrix}) into Eq.(\ref{eq:ixc}), we can compute IXC up to the second order in $A_{L}$ and $A_{R}$:
\begin{align}
\triangle E\approx \triangle E^{(0)}+\delta\triangle
E^{(1)}+\delta\triangle E^{(2)} \label{coupling-energy}
\end{align}
The first-order energy correction will be zero because it is proportional to the averages $\langle A_{L}\rangle $ and $\langle  A_{R}\rangle $
which are set to be zero by construction.
The major correction, therefore, comes from the second order perturbation and can be separated into correlation and uncorrelation terms:
\begin{equation}
\delta\triangle E^{(2)}=\delta\triangle E_{c}^{(2)}+\delta\triangle E_{uc}^{(2)}
\label{eq:e2}
\end{equation}
where
\begin{align}
\notag\frac{\delta\triangle E_{c}^{(2)}}{W}&=-2\mathrm{Im}\int_{-\infty}^{E_{F}} \frac{dE}{2\pi} \int\frac{dk_{y}}{2\pi}
\sum_{q_{y}}(q'_{Lx}-q_{x})\\
\notag&\times (k'_{Rx}-k_{x})\frac{t_{L;k_{x},k'_{Lx}}t_{R;q_{x},q_{Rx}}e^{i(q_{x}+k_{x})D}}{1-M(\vec{k})}\\
&\times \Big(1+\frac{M(\vec{q})+M(\vec{k})}{2-2M(\vec{q})}\Big)\langle k_{y}|A_{L}|q_{y}\rangle \langle q_{y}|A_{R}|k_{y}\rangle ,
\label{eq:c}\\
\notag\frac{\delta\triangle E_{uc}^{(2)}}{W}&=2\mathrm{Im}\sum_{j=L,R}\int_{-\infty}^{E_{F}} \frac{dE}{2\pi} \int\frac{dk_{y}}{2\pi}
2k_{x}\frac{M(\vec{k})}{1-M(\vec{k})}\\
\notag &\times \Big\{ k'_{jx}\langle A^{2}_{j}\rangle -\sum_{q_{y}}(q'_{jx}-q_{x})|\langle k_{y}|A_{j}|q_{y}\rangle |^{2}\\
&+\sum_{q_{y}}q_{x}\frac{M(\vec{q})}{1-M(\vec{q})}|\langle k_{y}|A_{j}|q_{y}\rangle |^{2}\Big\}
\label{eq:uc}
\end{align}
and
\begin{equation}
M(\vec{k})=r_{L;k_{x},k'_{Lx}}r_{R;k_{x},k'_{Rx}}e^{2ik_{x}D}.
\end{equation}
Note that these results are general, which will reduce to Eq.(3) in Ref.\cite{pre-art} when restricted to the double limits of $k_F \lambda_C \gg 1$ and $k_{F}D\gg 1$, i.e., a wide spacer.
The reason is that $k_{F}D\gg 1$ allows us to ignore the energy contribution from higher order
round-trip reflections in Eq.(\ref{eq:c}) and Eq.(\ref{eq:uc}):
\begin{align}
\notag\frac{\delta\triangle E_{c}^{(2)}}{W}&\approx -2\mathrm{Im}\int_{-\infty}^{E_{F}} \frac{dE}{2\pi} \int\frac{dk_{y}}{2\pi}
\sum_{q_{y}}(q'_{Lx}-q_{x})\\
\notag&\times (k'_{Rx}-k_{x})t_{L;k_{x},k'_{Rx}}t_{R;q_{x},q_{Rx}}e^{i(q_{x}+k_{x})D}\\
&\times\langle k_{y}|A_{L}|q_{y}\rangle \langle q_{y}|A_{R}|k_{y}\rangle
\label{eq:ca}\\
\notag\frac{\delta\triangle E_{uc}^{(2)}}{W}&\approx 2\mathrm{Im}\sum_{j=L,R}\int_{-\infty}^{E_{F}} \frac{dE}{2\pi} \int\frac{dk_{y}}{2\pi}
2k_{x}M(\vec{k})\\
&\times \Big\{ k'_{jx}\langle A^{2}_{j}\rangle -\sum_{q_{y}}(q'_{jx}-q_{x})|\langle k_{y}|A_{j}|q_{y}\rangle |^{2}\Big\}.
\label{eq:uca}
\end{align}
In the mean time,  $k_F \lambda_C \gg 1$ permits us to assume that the reflected/transmitted momenta $q/q'$ at the interface are not much different from their values without the roughness.
Then, only the $q_x\approx k_x$ survives in the brackets and can be pulled out of the summation to reproduce  Eq.(3) in Ref.\cite{pre-art}.

\section{Response function}
For a more systematic study of the IR in IXC,
we denote the Fourier component $\langle k_{y}|A_{j}|q_{y}\rangle $ by $H_{j}(q_{y}-k_{y})$ and all terms in Eq.(\ref{eq:c}) and Eq.(\ref{eq:uc})
are proportional to
\begin{equation}
\notag H_{j}(q_{y}-k_{y})H_{l}(k_{y}-q_{y}).
\end{equation}
For simplicity, assume that the two side layers are made of the same material and so the index $j$ may be omitted in the scattering coefficients.
Then, after changing the integration variable from $q_{y}$ into $P_{y}=q_{y}-k_{y}$, we can rewrite Eq.(\ref{eq:c}) as
\begin{equation}
\frac{\delta \triangle E_{c}^{(2)}}{W}=\sum_{P_{y}}G^{c}(P_{y})H_{L}(P_{y})H_{R}(-P_{y})
\label{eq:gc1}
\end{equation}
where
\begin{align}
\notag G^{c}(P_{y})&=-2\mathrm{Im}\int_{-\infty}^{E_{F}} \frac{dE}{2\pi} \int\frac{dk_{y}}{2\pi}4k_{x}q_{x}\\
&\times\frac{\sqrt{M(\vec{k})M(\vec{q})}}{1-M(\vec{k})}\Big(1+\frac{1}{2}\frac{M(\vec{q})+M(\vec{k})}{1-M(\vec{q})}\Big).
\label{eq:gc2}
\end{align}
Similarly,  Eq.(\ref{eq:uc}) becomes
\begin{equation}
\frac{\delta \triangle E_{uc}^{(2)}}{W}=\sum_{j,P_{y}}G^{uc}(P_{y})H_{j}(P_{y})H_{j}(-P_{y})
\label{eq:guc1}
\end{equation}
where
\begin{align}
\notag G^{uc}(P_{y})&=2\mathrm{Im}\int_{-\infty}^{E_{F}} \frac{dE}{2\pi} \int\frac{dk_{y}}{2\pi}2k_{x}\\
&\times\frac{M(\vec{k})}{1-M(\vec{k})}\Big(k'_{x}-q'_{x}+\frac{q_{x}}{1-M(\vec{q})}\Big)
\label{eq:guc2}
\end{align}
and
\begin{equation}
M(\vec{k})=r^{2}_{k_{x},k'_{x}}e^{2ik_{x}D}
\label{eq:mk}
\end{equation}
and the momenta $q_{x}$ and $q'_{x}$ are functions of $q_{y}=P_{y}+k_{y}$.
The $P_{y}$ in $G(P_{y})$ signifies the momentum transfer induced by a given Fourier component of the interface profile.
The response function $G(P_{y})$ is determined by the reflection coefficients, momenta and the exponential factors from the
round-trip propagations between interfaces.


The study in the Casimir effect concluded\cite{uncor-the} that the perturbation method would become equivalent to PFA in the limit of long corrugation wavelengths. We shall prove in the following that this statement remains true for IXC.
By use of the relation $G(P_{y})=G(-P_{y})$ implied by Eq.(\ref{eq:gc2}) and Eq.(\ref{eq:guc2}) and taking the limit $P_{y}\rightarrow 0$,
the sum of Eq.(\ref{eq:gc1}) and Eq.(\ref{eq:guc1}) becomes
\begin{align}
\notag\frac{\delta\triangle E^{(2)}}{W}&\approx\mathrm{Im}\int_{-\infty}^{E_{F}} \frac{dE}{2\pi} \int\frac{dk_{y}}{2\pi}4k^{2}_{x}\frac{M(\vec{k})}{(1-M(\vec{k}))^{2}}\\
&\times\sum_{P_{y}}\Big|H_{L}(P_{y})-H_{R}(P_{y})\Big|^{2}.
\label{pfa1}
\end{align}
where the summation can be carried out to give
\begin{equation}
\sum_{P_{y}}\Big|H_{L}(P_{y})-H_{R}(P_{y})\Big|^{2}=\big\langle (A_{L}(y)-A_{R}(y))^{2}\big\rangle .
\end{equation}

Summarizing the above calculations, IR introduces a shift to the coupling energy $\triangle E^{(0)}$ in Eq.(\ref{IXC}) for interfaces with mild corrugations:
\begin{align}
\frac{\delta\triangle E^{(2)}}{W}\approx\frac{1}{2}
\big\langle (A_{L}(y)-A_{R}(y))^{2}\big\rangle \frac{d^{2}(\triangle E^{(0)}/W)}{d D^{2}}
\label{pfa2}
\end{align}
under the limit of $A\ll 2\pi/k_F\ll \lambda_C$.
Note that Eq.(\ref{pfa2}) is of the form of PFA
which Taylor expands the variation $A_{L}(y)-A_{R}(y)$ to the second order for the coupling energy.
This demonstrates that Eq.(\ref{eq:gc1}) and Eq.(\ref{eq:guc1}) are more general than PFA since they do not require $k_F \lambda_C\gg 1$.
In the special case of $A_{L}(y)=A_{R}(y)\neq 0$, our method can still capture the effects of IR while PFA predicts none.

In the next section, we apply our method to real systems where $k_F \lambda_C$ is not necessarily large, and concentrate on the effect of correlation term Eq.(\ref{eq:gc1}) and Eq.(\ref{eq:gc2}). Discrepancies between our results and those of PFA will be highlighted.
The propagation term will be shown to display interesting features from the quantum interference for GMR. By increasing the potential barrier $V_0$ of the spacer above $E_F$, our previous results can be applied to TMR.
That turns the propagation term into a decaying function of the spacer width and the mathematical form of Eq.(\ref{eq:gc1}) through Eq.(\ref{eq:guc2}) shall bear more resemblance to those of the Casimir effect\cite{uncor-the,cor-the}.

\section{Correlated Interfaces}
Let us start from simple sinusoidal functions for the corrugation on two correlated interfaces:
\begin{align}
\notag &A_{L}(y)=a_{L}\cos(py), \\
&A_{R}(y)=a_{R}\cos(p(y+b))
\label{cor-plates}
\end{align}
where $p=2\pi/\lambda_{c}$. The energy correction thus depends on the lateral mismatch $b$.
Plugging Eq.(\ref{cor-plates}) into Eq.(\ref{eq:gc1}) gives
\begin{equation}
\frac{\delta\triangle E_{c}^{(2)}}{W}=\frac{a_{L}a_{R}}{2}\cos{(pb)}G^{c}(p).
\label{eq:cor-e}
\end{equation}
Similar to the procedures from Eq.(\ref{pfa1}) to Eq.(\ref{pfa2}), Eq.(\ref{eq:cor-e}) can be turned into the PFA form in the limit of $p\rightarrow 0$ and
\begin{align}
G^{c}(p\rightarrow 0)=-\frac{d^{2}(\triangle E^{(0)}/W)}{d D^{2}}.
\end{align}

Equation (\ref{eq:cor-e}) implies the correlation energy correction can be modulated by a sinusoidal function of the phase difference
between the two interfaces.
Since the uncorrelation term does not depend on the phase difference,
the response function $G^c (p)$ for the correlation term can be measured by substracting the coupling energy in Eq.(\ref{coupling-energy}) for in-phase case from that for out-of-phase.
In the next two subsections,  we shall use the sensitivity function
\begin{equation}
\rho^{c}(p)=\frac{G^{c}(p)}{G^{c}(0)}
\end{equation}
to quantify the discrepancy between our results and the  PFA ones.

\subsection{TMR}
\begin{figure}[h!]
\includegraphics[width=0.45\textwidth]{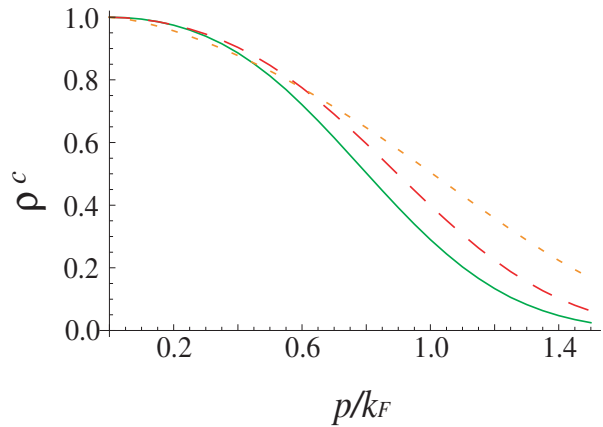}
\caption{(Color online) Sensitivity function $\rho^c$ is plotted as a function of $p/k_{F}$ for $V_{0}/E_{F}=2$ and $k_{F}D=3$ (in dotted line), $7$ (dashed line),
and $10$ (solid line).}
\label{correlation-decay}
\end{figure}

The sensitivity function $\rho^{c}$ for a typical TMR system is plotted in
Fig.{\ref{correlation-decay}} as a function of $p/k_{F}$ for different values of spacer width $D$.
The  $\rho^{c}$ in this figure exhibits three traits: (1) it never exceeds unity which implies that the PFA always overestimates the correlation effect for TMR;
(2) it approaches unity at small $p/k_{F}$ when our method reduces to the PFA; (3) it decays exponentially to zero when $p/k_{F}$ becomes large, which is corroborated by our analytic derivations for the asymptotic form of $G^{c}(p)=\alpha p\ e^{-pD}$ at $p/k_{F}\gg 1$ by Taylor expanding Eq.(\ref{eq:gc2}). The parameter $\alpha$ depends on $E_{F}$ and $V_{0}$.
These features are shared by the Casimir effect\cite{cor-the}
because of their similar mathematical formalism.

\subsection{GMR}
\begin{figure}[h!]
\includegraphics[width=0.45\textwidth]{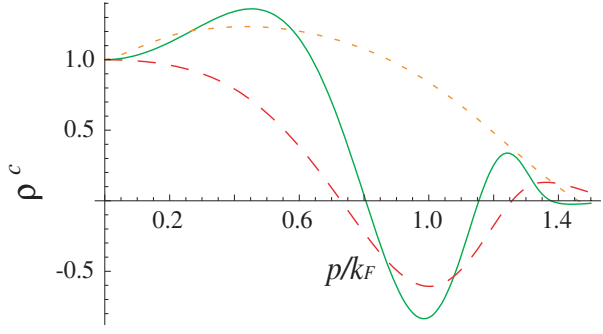}
\caption{(Color online) Variation of $\rho^c$ as a function of $p/k_{F}$ for $V_{0}/E_{F}=0.5$ and $k_{F}D=3$ (in dotted line), $7$ (dashed line),
and $10$ (solid line).}
\label{PFA-correlation}
\end{figure}

As we reduce the potential barrier to $V_{0}/E_{F}=0.5$, the system enters the GMR regime.
The results are plotted in Fig.{\ref{PFA-correlation}}.
Comparing to Fig.\ref{correlation-decay} for the TMR, the sensitivity function becomes oscillatory and can be negative in certain ranges of $p$.
The period of oscillation shortens as the spacer gets thicker
because the spacer width $D$ is multiplied to the corrugation period $p$ in the phase term. Furthermore, $\rho^c$ can now exceeds unity which is a necessary condition for the significant enhancement of IXC by IR\cite{pre-art}.
Detail derivations for the enhancement and this statement are arranged in the following section.

\begin{figure}[h!]
\includegraphics[width=0.4\textwidth]{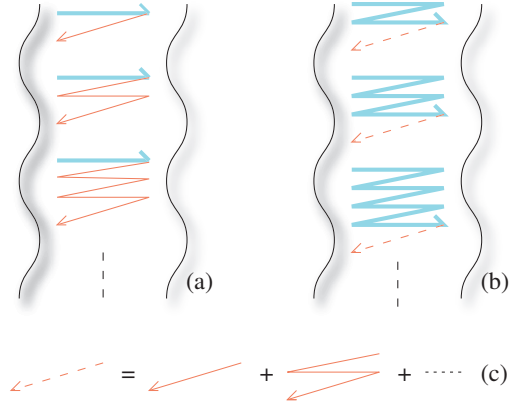}
\caption{(Color online) Diagrams for the scattering paths in (a) the first and (b) second terms of the response function in Eq.(\ref{eq:gc21}) and Eq.(\ref{eq:gc22}), respectively.
The thick/thin lines denote the paths with momenta $\vec{q}/\vec{k}$.
(c) The dashed line denotes the path which is contributed by the thin line with any number of loops }
\label{diagram}
\end{figure}

\begin{figure}[h!]
\includegraphics[width=0.45\textwidth]{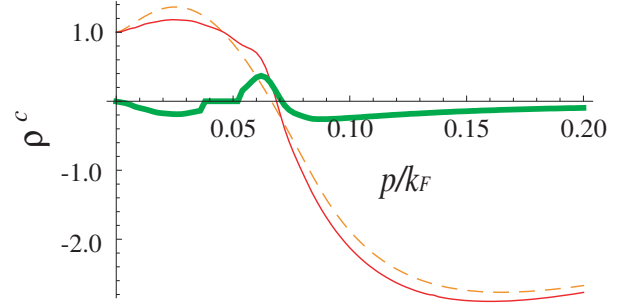}
\caption{(Color online) Variation of $\rho^c$ versus $b/k_{F}$ with $V_{0}/E_{F}=-0.2$, $k_{F}D=3$ and for Eq.(\ref{eq:gc21}) (dashed line),
Eq.(\ref{eq:gc22}) (thick line) and their sum (thin line).}
\label{correlation-fano}
\end{figure}

If the spacer potential is changed into a well, the emerging bound states is expected to have a limiting role at mediating IXC since their probability decay exponentially into the metallic side layers. This is indeed true for smooth interfaces.
However, in the presence of IR, the new eigenstates are a mixture of both scattering and bound states. And, according to Eq.(\ref{eq:mk}), the bound states can contribute and render the denominator, $1-M(\vec{q})$, in Eq.(\ref{eq:gc22}) vanishing. So it is expected to generate new features in IXC.
We divide the response function at Eq.(\ref{eq:gc2}) into two terms:
\begin{align}
G_{(1)}^{c}(P_{y})&=-2\mathrm{Im}\int_{-\infty}^{E_{F}} \frac{dE}{2\pi} \int\frac{dk_{y}}{2\pi}4k_{x}q_{x}\frac{\sqrt{M(\vec{k})M(\vec{q})}}{1-M(\vec{k})}
\label{eq:gc21}\\
\notag G_{(2)}^{c}(P_{y})&=-2\mathrm{Im}\int_{-\infty}^{E_{F}} \frac{dE}{2\pi} \int\frac{dk_{y}}{2\pi}4k_{x}q_{x}\frac{\sqrt{M(\vec{k})M(\vec{q})}}{1-M(\vec{k})}\\
&\times\frac{1}{2}\frac{M(\vec{q})+M(\vec{k})}{1-M(\vec{q})}.
\label{eq:gc22}
\end{align}
The first term is plotted in Fig.\ref{diagram}(a), which represents the contribution from
the paths that are scattered to momentum  $\vec{q}$ in the beginning and then reflected to the original momentum $\vec{k}$.
The second term, as shown in Fig.\ref{diagram}(b) and is directly related to the additional resonance transmission,
represents the contribution from the remaining paths that are transmitted in momentum $\vec{q}$ for more than one loop before
being reflected to $\vec{k}$.
To clarify the dissimilarity of these two terms, we calculated a well system in Fig.\ref{correlation-fano} and show
the magnitudes of the first term, second term, and the total value of the response function.
Again, the response function is divided by $G{c}(0)$ in the figure.
Because the well system for a trilayer exhibits two bound states which generate
two singularities in the integral of Eq.(\ref{eq:gc22}) for resonance transmissions,
the second term of the response function displays two kinks as we modulate the corrugation period.

Although we use a perturbative method for the calculation of IXC energy between rough interfaces,
the result of correction in GMR system still present unusual characteristics from the quantum interference and the resonance states.
The PFA is not relevant to this regime.

\section{Enhancement of IXC}
Equation (\ref{coupling-energy}) consists of two parts, correlation term in Eq.(15) and uncorrelated one in Eq.(16). By use of Eq.(\ref{pfa2}) from PFA, IXC for mildly corrugated interfaces can be expressed as:
\begin{align}\label{eq:c+uc}
\notag\Delta E&\approx\Delta E^{(0)}-\big\langle A_{L}(y)A_{R}(y)\big\rangle \frac{d^{2}\triangle E^{(0)}}{d D^{2}}\\
&+\frac{1}{2}\big\langle A_{L}^{2}(y)+A_{R}^{2}(y)\big\rangle \frac{d^{2}\triangle E^{(0)}}{d D^{2}},
\end{align}
where the $A_{L/R}$ and $D$ are the same definitions as in previous sections. 
Since $\Delta E^{(0)}$ for GMR is an oscillatory function of the spacer width $D$, it shares the same sign as the negative of its second derivative for $k_{F}D>1$. As a result, the second term always strengthens IXC, while the third term diminishes it.
An overall enhancement of the coupling strength is realized when the correlation term dominates. However, this is not possible in the above PFA expression for mild corrugations. The sum of these two terms can never be positive and, at most, cancel each other to give null contribution when the tomography on both interfaces happen to be unrealistically identical. Therefore, it is safe to say that the IR also suppresses IXC within the second-order perturbation of PFA. 

\begin{figure}[h!]
\includegraphics[width=0.45\textwidth]{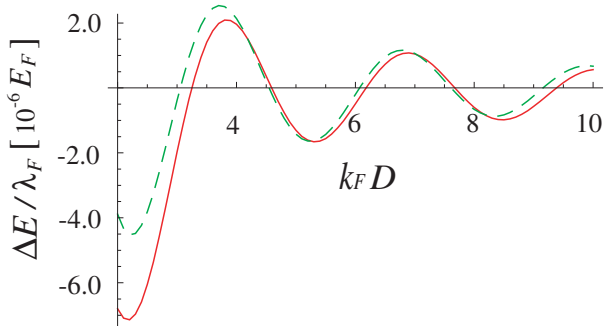}
\caption{(Color online) Coupling strength is plotted as a function of $k_{F}D$ for two interfaces that share the same topography $A(y)=1/(2k_{F}) \sin(k_{F}y/2)$ with $V_{0}/E_{F}=-0.2$. The result for smooth interfaces is shown in the dashed line for comparison. }
\label{deltae-enhancement}
\end{figure}

We shall now demonstrate that more severe corrugations, $A\ll 2\pi/k_{F} \sim\lambda_{c}$, and correlated tomographies are two essential ingredients to enhance IXC.  The former requires us to improve upon the PFA within the second-order perturbation, while the latter brings in strong quantum interference.
To clarify this statement, we use our method to estimate the coupling strength for the 2D trilayer system with $V_{0}/E_{F}=-0.2$ and identical topography $A(y)=1/(2k_{F}) \sin(k_{F}y/2)$ on both interfaces. As shown in Fig.\ref{deltae-enhancement}, the effect of correlated and severe corrugations can improve IXC for smooth interfaces by as much as one and a half times.

\begin{figure}[h!]
\includegraphics[width=0.45\textwidth]{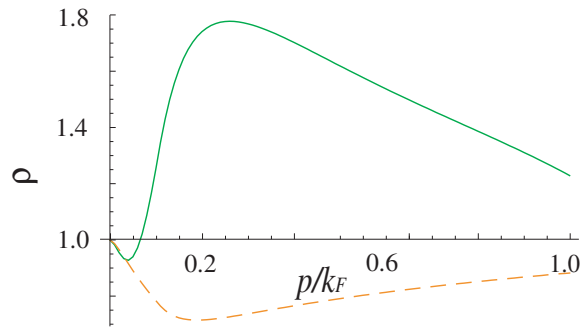}
\caption{(Color online) Setting the corrugation function $A(y)$ to observe the sinusoidal form, $\sin py$, the correlation $\rho^{c}$ (solid line) and uncorrelation sensitivity functions $\rho^{uc}$ (dashed line) are plotted as a function of  $p/k_{F}$ for $V_{0}/E_{F}=-0.2$ and $k_{F}D=2.2$.}
\label{c+uc-enhancement}
\end{figure}

It is heuristic to approximate $d^{2}\triangle E^{(0)}/d D^{2}$ by $-4k^{2}_{F}E^{(0)}$ when $k_{F}D\gg1$ since IXC for two identical interfaces with topography $A(y)$ can be neatly reduced to
\begin{align}\label{eq:approxx}
\Delta E(p)&\approx\Delta E^{(0)}\Big[1+4\big(\rho^{c}-\rho^{uc}\big)k_{F}^{2}\big\langle A(y)^{2}\big\rangle\Big].
\end{align}
It is then clear that an enhancement in IXC is being caused by the dominance of the correlation sensitivity function over the uncorrelation one, which can be realized for a wide range of $p/k_F$ in Fig.\ref{c+uc-enhancement}. The largest enhancement appears around $p=0.22k_{F}$, which is about two times that for smooth interfaces.
We checked that Eq.(\ref{eq:approxx}) gave roughly the same value as that without the approximation, 1.4 and 1.5 respectively for the topography and parameter in Fig.\ref{deltae-enhancement}.

We have demonstrated that our approach is more general than the PFA at capturing the effect of quantum interference among different reflected paths of carriers within the spacer. This allows us to obtain the enhancement of IXC when $p$ is comparable to $k_F$. It is then reasonable to ask what happens when the characteristic corrugation $p$ is much greater than $k_{F}$. It turns out that the momentum change along the surface of the interfaces, being of order $p$, is so large that the longitudinal momentum becomes pure imaginary during elastic collisions. This means that these paths only survive a short distance in the $x$-direction and, therefore, are not expected to lead to major interference. 
To be more rigorous, if we approximate the additional scattering momenta ${q_{x},q'_{x}}$ due to IR by $ip$ and insert into Eq.(21) and Eq.(23), the sensitivity function for correlation term will decay as $p\ e^{-pD}$ while the uncorrelation one remains roughly a constant. 
In retrospect, the process of allowing different reflected paths of carriers to interfere is similar to that of localization, but the eventual effect is different. In this case of IXC, although the second term in Eq.(\ref{eq:approxx}) contains the second power of the small perturbation parameter $k_F A$, the quantum interference is still capable of rendering this term large via the other factor, $\rho^{c}-\rho^{uc}$. 

\section{Discussions and Conclusions}
Our perturbative approach to evaluate the effects of interface roughness in the trilayers  was motivated by a similar effort in the Casimir problem. 
To be precise, the role of interface roughness at causing interference among reflected electromagnetic waves  within the cavity finds a better analogy in TMR, rather than GMR. Reason being that the decaying nature of carrier wavefunctions in TMR limits their quantum interference within the spacer, while the fact that virtual photons lack a characteristic length scale like the inverse of Fermi momentum $1/k_F$ dilutes any possible constructive interference after all wavelengths are summed over. 
In contrast, the quantum interference survives and has a nonnegligible effect on GMR when the characteristic wavelength of corrugations is shared by both interfaces and comparable to $1/k_F$.

Compared to our previous study\cite{pre-art}, a couple of improvements have been made in this work. First, although both calculations retained up to the second order in the corrugation amplitude $A(y)$, we included more loops of multiple scattering from the smooth part of the interfaces. Furthermore, $A(y)$ was no longer confined to be of the sinusoidal form. Second, an analogy to the Casimir effect was made, which allowed us to borrow the concept of response and sensitivity functions as indicators of the extent of influence by $A(y)$ without having to know its detailed form. We followed up by more discussions on similar and different effects of  $A(y)$ in trilayers and the Casimir mirrors. Third, more thorough derivations were done to compare our approach with the prevailing proximity-force approximation for different periods of corrugations. This enabled a better quantitative estimate of the enhancement from quantum interference. 

In conclusion, 
we find that the perturbative approach reaches the same conclusions as the proximity-force approximation in the limit of $p\ll k_{F}$. Namely, mild corrugations lead to a suppression of the interference and thus the interlayer exchange coupling.
Correlated roughness with short wavelengths gives rise to several interesting features: (1) The energy correction oscillates as we vary the corrugation wavelength. (2) The magnitude of
correction can be larger than the prediction made by the proximity-force approximation. 
(3) While they are expected by the proximity-force approximation to be irrelevant to the transmission coefficient, the bound states within the spacers are found to affect the resonance 
transmission through several kinks in the energy correction.
One last important feature concerns the enhancement of interlayer exchange coupling by the interface roughness. Its occurrence relies on further requirement that the Fourier conjugates alluded to above be close to the Fermi momentum.

Support by the National Science Council in Taiwan under Grant No. 98-2112-M007-005-MY3 is acknowledged.

\begin{appendix}
\section*{Appendix}
\section{Connection between The Casimir effect and IXC}
In this appendix, we would like to extend the concept of radiation pressure in the Casimir effect to the IXC problem.
We start from the Casimir energy which is the summation of zero-point energies for quantum states in the presence of boundaries:
\begin{align}
E=\frac{1}{2}\sum_{n} \omega_{n}
\label{eq:zero-point}
\end{align}
where $n$ and $\omega_{n}$ denote the $n$-th bound state with frequency $\omega_{n}$.
In contrast, the IXC energy measures the increase in the total carrier energy when considering the boundaries:
\begin{align}
\Delta E=\sum^{N}_{n} E_{n}-E_{r}
\label{eq:couple-E}
\end{align}
where $N$ is total number of carriers and $E_{r}$ is the reference energy without the boundary.

\subsection{Radiation Force in One Dimension}

To clarity the connection between these two energies, we now start from the radiation force of fields in a one-dimensional potential well with width $D$.
The eigenstates of carriers in a quantum well can be written as a combination
of two travelling waves in opposite directions:
\begin{align}
\sqrt{\frac{2}{D}}\sin(k_{n}x)=-i\sqrt{\frac{1}{2D}}\Big[ e^{ik_{n}x}-e^{-ik_{n}x}\Big] =\overrightarrow{e_{n}}_{\mathrm{C}}+\overleftarrow{e_{n}}_{\mathrm{C}}
\label{eq:cavity-state}
\end{align}
where $k_{n}={n\pi}/{D}$ is the quantized momentum.

The radiation force measures the impetus per unit time contributed by the carriers in the cavity:
\begin{align}
\notag \Delta F(D)&=\sum_{n}^{N}\frac{\Delta p_{n}}{\Delta t_{n}}\big\langle\overrightarrow{e_{n}}_{\mathrm{C}}\cdot \overrightarrow{e_{n}}_{\mathrm{C}}^{\dag}
+\overleftarrow{e_{n}}_{\mathrm{C}}\cdot \overleftarrow{e_{n}}_{\mathrm{C}}^{\dag} \big\rangle\\
\notag&=\sum_{n}^{N}\frac{2 k_{n}}{2D/v_{n}}\big\langle\overrightarrow{e_{n}}_{\mathrm{C}}\cdot \overrightarrow{e_{n}}_{\mathrm{C}}^{\dag}
+\overleftarrow{e_{n}}_{\mathrm{C}}\cdot \overleftarrow{e_{n}}_{\mathrm{C}}^{\dag} \big\rangle\\
&=\sum_{n}^{N}\frac{2E_{n}}{D} \big\langle\overrightarrow{e_{n}}_{\mathrm{C}}\cdot \overrightarrow{e_{n}}_{\mathrm{C}}^{\dag}
+\overleftarrow{e_{n}}_{\mathrm{C}}\cdot \overleftarrow{e_{n}}_{\mathrm{C}}^{\dag} \big\rangle
\label{eq:ixc-1d-force}
\end{align}
where $v_{n}= k_{n}/m$ and $m$ denotes the carrier mass.
The IXC energy can be evaluated by
\begin{align}
\notag\Delta E&=-\int^{D}_{\infty} dD'\Delta F(D')\\
&=\sum_{n}^{N}E_{n}\big\langle\overrightarrow{e_{n}}_{\mathrm{C}}\cdot \overrightarrow{e_{n}}_{\mathrm{C}}^{\dag}
+\overleftarrow{e_{n}}_{\mathrm{C}}\cdot \overleftarrow{e_{n}}_{\mathrm{C}}^{\dag} \big\rangle -E_{r}.
\end{align}
It is not surprising that the above expression reduces to Eq.(\ref{eq:couple-E}) upon assigning
$\langle\overrightarrow{e_{n}}_{\mathrm{C}}\cdot \overrightarrow{e_{n}}_{\mathrm{C}}^{\dag}\rangle
=\langle\overleftarrow{e_{n}}_{\mathrm{C}}\cdot \overleftarrow{e_{n}}_{\mathrm{C}}^{\dag} \rangle=1/2$.

Similar derivations for the vaccuum states give:
\begin{equation}
F=\sum_{n}\frac{\omega_{n}}{D} \big\langle\overrightarrow{e_{n}}_{\mathrm{C}}\cdot \overrightarrow{e_{n}}_{\mathrm{C}}^{\dag}
+\overleftarrow{e_{n}}_{\mathrm{C}}\cdot \overleftarrow{e_{n}}_{\mathrm{C}}^{\dag} \big\rangle_{\mathrm{vac}}
\end{equation}
and
\begin{equation}\label{eq:vac}
E=\sum_{n} \omega_{n}\big\langle\overrightarrow{e_{n}}_{\mathrm{C}}\cdot \overrightarrow{e_{n}}_{\mathrm{C}}^{\dag}
+\overleftarrow{e_{n}}_{\mathrm{C}}\cdot \overleftarrow{e_{n}}_{\mathrm{C}}^{\dag} \big\rangle_{\mathrm{vac}}=\frac{1}{2}\sum_{n} \omega_{n}.
\end{equation}
where the extra coefficient $1/2$ comes from the fact that the quantum amplitude for the vaccuum state is just one half of
the corresponding commutator from photon operators\cite{lossy-cavity-casimir}.

\subsection{Radiation Force in Three Dimensions}

\begin{figure}
\includegraphics[width=0.5\textwidth]{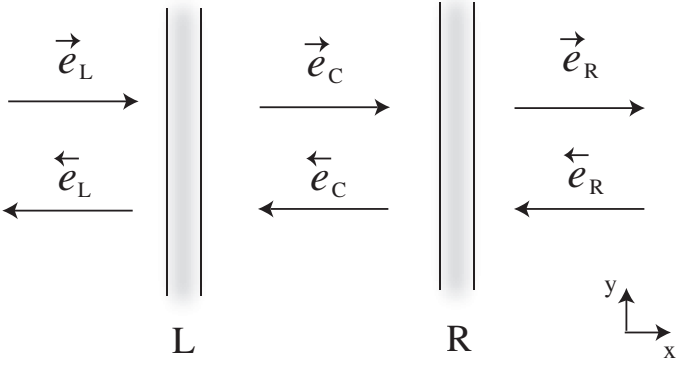}
\caption{Schematic plot of a trilayer system to clarify our notations: $L/R$ denote the fields in the left/right sides of the spacer, whereas $C$ represents those in the spacer.}
\label{pressure}
\end{figure}

We generalize the result in Eq.(\ref{eq:ixc-1d-force}) to estimate the radiation force in a 3D trilayer system in Fig.\ref{pressure}.
\begin{align}
\notag\Delta F_{L}(D)=&\sum_{k}^{k_{F}}\frac{2E_{k}}{L_{x}} \cos^{2} \theta\big\langle\overrightarrow{e_{k}}_{\mathrm{C}}\cdot \overrightarrow{e_{k}}_{\mathrm{C}}^{\dag}
+\overleftarrow{e_{k}}_{\mathrm{C}}\cdot \overleftarrow{e_{k}}_{\mathrm{C}}^{\dag}\\
\notag &-\overrightarrow{e_{k}}_{\mathrm{L}}\cdot \overrightarrow{e_{k}}_{\mathrm{L}}^{\dag}
-\overleftarrow{e_{k}}_{\mathrm{L}}\cdot \overleftarrow{e_{k}}_{\mathrm{L}}^{\dag}\big\rangle\\
=&-\Delta F_{R}(D)
\label{eq:quantum-force}
\end{align}
where $L/R$ denote the fields in the left/right sides of the trilayer and $L_{x}$ is the system length in $x$ direction.
The extra factor $\cos^{2} \theta$
comes from projections of the momenta and velocities on the normal direction of interfaces.
Take the continuum limit and the summation can be changed to an integral
\begin{align}
\notag\sum_{k}^{k_{F}}=&A\int\frac{d^{2}k_{\|}}{(2\pi)^{2}}L_{x}\frac{dk_{x}}{2\pi}\\
=&A\int_{\mathrm{IBZ}}\frac{d^{2}k_{\|}}{(2\pi)^{2}}L_{x}\int^{\mathrm{E_{F}}}_{0}\frac{dE}{2\pi}\frac{m}{k_{x}}.
\end{align}
where  $A=L_{y}L_{z}$ and the notation
IBZ signalizes the range of integral to be bounded by the interfacial Brillouin zone.
The amplitude of quantum fields can be obtained from the scattering states in the trilayer system:
\begin{align}
\notag &\psi_{k,>}(\vec{r})= \left\{\begin{array}{ll}
e^{ik_{x}x+i\vec{k_{\|}}\cdot\vec{r_{\|}}}+R_{>}e^{-ik_{x}x+i\vec{k_{\|}}\cdot\vec{r_{\|}}}, &\mathrm{left-side} \\
C_{>}e^{ik_{x}x+i\vec{k_{\|}}\cdot\vec{r_{\|}}}+D_{>}e^{-ik_{x}x+i\vec{k_{\|}}\cdot\vec{r_{\|}}}, &\mathrm{spacer}\\
T_{>}e^{ik_{x}x+i\vec{k_{\|}}\cdot\vec{r_{\|}}},   &\mathrm{right-side}
\end{array} \right.\\
\label{scattering-states}
\end{align}
where $ >$ denotes the moving direction of the scattering state. The left-moving one $\psi_{k,<}$ can be defined similarly.
Since these two states are orthogonal,
their contributions to the inner product of field amplitudes can be separated:
\begin{align}
\langle\overrightarrow{e_{k}}_{\mathrm{C}}\cdot \overrightarrow{e_{k}}_{\mathrm{C}}^{\dag}\rangle
= \sum_{\phi=>,<} |C_{\phi}|^{2}
\end{align}

Based on the above discussions and the relations $T_{<}=T_{>}$, $|R_{>}|^{2}+|T_{>}|^{2}=1$,
the quantum radiation force in Eq.(\ref{eq:quantum-force}) can be rearranged as
\begin{align}
\frac{\Delta F_{L}(D)}{A}
\notag&= \int^{\mathrm{E_{F}}}_{0}\frac{dE}{2\pi}\int_{\mathrm{IBZ}}\frac{d^{2}k_{\|}}{(2\pi)^{2}}k_{x}\\
&\times\Big[\sum_{\phi=>,<}\big\langle |C_{\phi}|^{2}+|D_{\phi}|^{2}\big\rangle-2 \Big]
\label{eq:quantum-force2}
\end{align}
where $|C_{\phi}|^{2}$ and $|D_{\phi}|^{2}$ are functions of $\{T_{L},T_{R},R_{L},R_{R}\}$
which denote the transmission and reflection coefficients at each of the barriers in Fig.\ref{pressure}.
Equation (\ref{eq:quantum-force2}) will become
\begin{align}
\notag\frac{\Delta F_{L}(D)}{A}&= -\int^{\mathrm{E_{F}}}_{0}\frac{dE}{2\pi}\int_{\mathrm{IBZ}}\frac{d^{2}k_{\|}}{(2\pi)^{2}}k_{x}
\Big[ 2\\
\notag&-\frac{|T_{L}|^{2}+|R_{L}T_{R}e^{ik_{x}D}|^{2}+|T_{R}|^{2}+|R_{R}T_{L}e^{ik_{x}D}|^{2}}{|1-R_{L}R_{R}e^{2ik_{x}D}|^{2}} \Big]\\
&=4\mathrm{Re}\int^{\mathrm{E_{F}}}_{0}\frac{dE}{2\pi}\int_{\mathrm{IBZ}}\frac{d^{2}k_{\|}}{(2\pi)^{2}}
k_{x}\frac{R_{L}R_{R}e^{2ik_{x}D}}{1-R_{L}R_{R}e^{2ik_{x}D}}.
\label{eq:quantum-forcef}
\end{align}
We can also calculate the IXC energy:
\begin{align}
\frac{\Delta E}{A}=2\mathrm{Im}\int^{\mathrm{E_{F}}}_{0}\frac{dE}{2\pi}\int_{\mathrm{IBZ}}\frac{d^{2}k_{\|}}{(2\pi)^{2}}\mathrm{ln}\big[1-R_{L}R_{R}e^{2ik_{x}D}\big]
\label{eq:pressure-ixc}
\end{align}
Equation (\ref{eq:pressure-ixc}), which has been proved in Ref.\cite{Bruno} by using the concept of
quantum interference and Green's function, is commonly used in the study of IXC.

In the Casimir problem, we follow the same procedures for the vaccuum state in Eq.(\ref{eq:vac})
to obtain
\begin{align}
\frac{F_{L}}{A}&=2\mathrm{Re}\sum_{p}\int_{0}^{\infty}\frac{d\omega}{2\pi}\int\frac{d^{2}k_{\|}}{(2\pi)^{2}} k_{x}
\frac{R_{L}R_{R}e^{2ik_{x}D}}{1-R_{L}R_{R}e^{2ik_{x}D}}
\label{eq:cisimir-force}\\
\frac{E}{A}&=\mathrm{Im}\sum_{p}\int_{0}^{\infty}\frac{d\omega}{2\pi}\int\frac{d^{2}k_{\|}}{(2\pi)^{2}}\mathrm{ln}\big[1-R_{L}R_{R}e^{2ik_{x}D}\big]
\label{eq:casimir-e}
\end{align}
where $p$ denotes the transverse electric and magnetic modes.
By use of the Cauchy theorem, we can shift the integration to the imaginary frequency axis and rewrite the Casimir force in Eq.(\ref{eq:cisimir-force}) and energy in Eq.(\ref{eq:casimir-e}) as
\begin{align}
\frac{F_{L}}{A}&=2\sum_{p}\int_{0}^{\infty}\frac{d\xi}{2\pi}\int\frac{d^{2}k_{\|}}{(2\pi)^{2}} \kappa
\frac{R_{L}R_{R}e^{-2 \kappa D}}{1-R_{L}R_{R}e^{-2 \kappa D}}
\label{eq:cisimir-force-f}\\
\frac{E}{A}&=\sum_{p}\int_{0}^{\infty}\frac{d\xi}{2\pi}\int\frac{d^{2}k_{\|}}{(2\pi)^{2}}\mathrm{ln}\big[1-R_{L}R_{R}e^{-2 \kappa D}\big]
\label{eq:casimir-e-f}
\end{align}
where $\kappa=\sqrt{k^{2}_{\|}+\xi^{2}/c^{2}}$.
Equation (\ref{eq:cisimir-force-f}) and Eq.(\ref{eq:casimir-e-f}) are the familiar formulae for lossy optical cavities.
A previous article \cite{lossy-cavity-casimir} has used the vaccuum radiation pressure to derive Eq.(\ref{eq:cisimir-force-f}).
Although similar in concepts to theirs, our derivations to relate the IXC and the Casimir energy are more straightforward.
One qualitative difference is that the upper bound of energy integration in IXC is bound by the Fermi energy which energy scale eventually renders the
IXC force oscillatory in $2k_{F}D$.

\end{appendix}

\end{document}